\documentclass[trackchanges,twocolumn,longbib]{aastex7}

\definecolor{c_teal}{HTML}{006060}
\definecolor{c_lightteal}{HTML}{edf2f2}
\definecolor{c_oraneg}{HTML}{CC8400}
\hypersetup{linkcolor=teal,
            citecolor=teal,
            filecolor=teal,
            urlcolor=teal,
            linktoc=page,
            pdfborder={0 0 0}}

\usepackage{float, graphicx, multirow}
\usepackage{adjustbox, booktabs, csquotes}

\usepackage{enumitem, url}
\usepackage[british]{babel}

\usepackage{amssymb, amsmath, wasysym, chemformula, chemfig, xfrac}
\usepackage[group-digits = none]{siunitx}
\usepackage[e]{esvect}

\newcommand{\hPa}{\hecto\Pa}
\newcommand{\radflux}{\W\per\m\squared}

\newcommand{\bs}[1]{\vec{#1}}
\newcommand{\dff}{\textrm{d}}

\shorttitle{Hydrological Cycle of Tidally-Locked Planets}
\shortauthors{D. A. Williams et al.}
\graphicspath{{./}{plots/}}

\begin{document}

\title{The Hydrological Cycle and Circulation of Tidally-Locked Planets with a Finite Moisture Inventory}

\author[orcid=0000-0002-5840-2411]{Daniel A. Williams}
\affiliation{Department of Mathematics and Statistics, University of Exeter, EX4 4QF, UK}
\email[show]{d.a.williams3@exeter.ac.uk}  

\author[orcid=0000-0002-5971-8995]{Geoffrey K. Vallis}
\affiliation{Department of Mathematics and Statistics, University of Exeter, EX4 4QF, UK}
\email[]{g.vallis@exeter.ac.uk}  

\author[orcid=0000-0002-4775-3259]{Stephen I. Thomson}
\affiliation{Department of Mathematics and Statistics, University of Exeter, EX4 4QF, UK}
\email[]{s.i.thomson@exeter.ac.uk}  

\author[orcid=0000-0003-1622-0545]{William J. M. Seviour}
\affiliation{Department of Mathematics and Statistics, University of Exeter, EX4 4QF, UK}
\email[]{w.seviour@exeter.ac.uk}

\begin{abstract}
Tidally-locked planets express a range of climate regimes distinct from those found in our solar system. Their atmospheres are often dominated by planetary-wide overturning circulations and super-rotating jets. In this paper we explore the hydrological cycle of such planets and how it might depend on the total inventory of condensate. To this end we use a set of idealised GCM simulations to probe the response of the hydrological cycle to the surface moisture budget in their atmospheres. We find that for low moisture inventories the climate quickly develops a barren day side and a moist night side where the moisture is trapped, as might be expected. For higher moisture inventories the drying of the day side is significantly delayed by the development of a substellar lake just east of the substellar point, which is sustained by the presence of a local thunderstorm-like convective system that persists even after the lake disappears. The timescales for the lake to disappear are millennia, suggesting that even a weak transfer of moisture through surface processes could maintain a moist day side.  We also find that adjusting the threshold at which the evaporative rate becomes constrained by the remaining surface liquid depth modifies the longevity of the lake and surrounding day side conditions. These findings have implications for understanding a diverse range of planetary climates and habitability, given the distribution of moisture and presence of a substellar lake.
\end{abstract}


\section{Introduction}
Every planet in our solar system experiences a diurnal radiation cycle, resulting from a difference between its axial and orbital rotation rate. As more exoplanets are discovered, evidence suggests that many of the confirmed low-mass, terrestrial planets are likely to be tidally-locked \citep{nasaexoplanetscienceinstitute2020-Planetary} as a result of the close orbit around their host star \citep{witte2002-Orbital,barnes2017-Tidal,heller2011-Tidal}. These tidally-locked planets lack a diurnal radiation cycle, with one hemisphere permanently facing the star and the other facing away into deep space. As such, their climates are distinct from any found in our solar system \citep{kaspi2015-ATMOSPHERIC, haqq-misra2018-Demarcating, komacek2019-Atmospheric}.

Tidally-locked climates have been investigated in various modelling studies, particularly regarding large-scale atmospheric circulation \citep{showman2002-Atmospheric, merlis2010-Atmospheric, pierrehumbert2019-Atmospheric}. More focused studies also exist, including of atmospheric wave dynamics \citep{showman2011-EQUATORIAL, koll2016-TEMPERATURE, hammond2018-Wavemean}, convection \citep{sergeev2020-Atmospheric} and heat transport \citep{noda2017-Circulation, lewis2022-Temperature}. Many dynamical studies studies have limited themselves to dry atmospheres \citep{joshi1997-Simulations, edson2011-Atmospheric} although others, often with a focus on habitability, have included moisture \citep{leconte2013-3D, yang2019-Simulations}. Direct observations of tidally-locked exoplanets are generally beyond current observational capabilities, so most studies are model-focussed, although some make a connection with observations including phase-curve data \citep{koll2015-DECIPHERING, penn2017-Thermal, penn2018-Atmospheric, song2021-Asymmetry}.

Of the studies that consider moist tidally-locked terrestrial planets, the majority use an aquaplanet configuration; that is, a water-covered planet with an infinite supply of moisture.  \citet{merlis2010-Atmospheric, edson2011-Atmospheric} and \citet{noda2017-Circulation} provide typical setups, although dynamical oceans are sometimes used \citep{salazar2020-Effect}. Aquaplanets are appropriate as they remove complications arising from topography and other land-surface interactions. The intermediate state between a dry planet and an aquaplanet, where the atmosphere is moist but land surface is present, is less frequently explored. For such tidally-locked planets, large surface temperature gradients are anticipated \citep{checlair2017-No, way2017-Resolving, lewis2018-Influence} along with strong large-scale atmospheric transport \citet{wordsworth2022-Atmospheres}, leading to the two hemispheres having substantially different conditions \citep{pierrehumbert2010-Principles}. Warmer tidally-locked climates have typically been explored to a lesser degree than colder ones, though GCM studies \citep{leconte2013-3D} exist, including of the hydrological cycle \citep{ding2020-Stabilization, labonte2020-Sensitivity} through changes to insolation and atmospheric composition.

% Theories of planetary evolution tell us that planets do not form in a tidally-locked state, but transition into one via orbital changes and angular momentum transfers \citep{barnes2017-Tidal}. Climate systems can therefore transition into a tidally-locked state, with associated reorganisation of atmospheric dynamics and distribution of any moisture inventory. 

% What can we predict about the hydrological cycle when the initial moisture inventory is varied: 
%given an initial equal distribution of surface moisture, how does 
In this paper we explore how the hydrological cycle evolves and is maintained when exposed to tidally-locked insolation. In particular, we explore how the initial inventory of water determines the final climatological state, using an idealised climate models to study the evolution of the atmospheric and surface state. The model configurations and experimental setup are introduced in Section 2. Results follow in Section 3, focussing in turn on the long-term climatology (Section 3.1), the evolution of the substellar region (Section 3.2); and the evolution of the surface hydrology (Section 3.3). Conclusions and comparisons with previous studies, along with a future outlook, follow in Section 4.

\section{Methods}
\subsection{Idealised GCM Setup}
Our tool of choice for this investigation is \textit{Isca} \citep{vallis2018-Isca}, a framework for constructing a suite of idealised climate models. \textit{Isca} allows for the construction of a range of model configurations and complexities \citep{thomson2019-Hierarchical}, making it particularly suited to modelling the atmospheres of Earth-like tidally-locked planets. We use \textit{Isca} to construct a basic planetary state with Earth-like radius and atmospheric composition, but that is tidally-locked. Specifically, the rotation period is set to 10 Earth days (0.1$\Omega_E$) and a tidally-locked insolation profile is used to ensure one hemisphere remains constantly illuminated. Obliquity and eccentricity are also set to zero to remove any seasonal cycle.

A spectral resolution of \texttt{T21} is used (32 latitude and 64 longitude values), giving a grid spacing of approximately \ang{5.8}$\times$\ang{5.6}. This relatively low resolution is used allows for very long integration times required by the experiments; the slow planetary rotation rate gives rise a large deformation radius and scales of motion that are reasonably well resolved at this resolution. The vertical grid is defined by 40 unevenly spaced $\sigma$-levels up to a height of \SI{0.15}{\hecto\Pa}, where $\sigma = p/p_s$ and $p_s$ is the surface pressure.

We use \textsc{Socrates} \citep{edwards1996-Studies, manners2016-SOCRATES} for the radiation scheme: a comprehensive, multi-band scheme developed by the UK Met Office which has already been used in applications beyond present-day Earth \citep{amundsen2016-UK, way2017-Resolving}. We run \textsc{Socrates} with the included \texttt{ga7} spectral files, providing 9 long-wave and 6 short-wave bands (standard configuration for an Earth-like model).%, and making use of stratospheric ozone absorption to provide an approximate stratospheric temperature structure.

\subsection{Hydrological Cycle}\label{sec:modified_buckets}
An aquaplanet cannot properly represent a true hydrological cycle since the ocean is assumed to be an infinite source of liquid and the surface cannot dry out. Here, we avoid this limitation in a relatively simple way by setting the surface to be `land' globally but using a `bucket' hydrology scheme. In a bucket model the surface within each model grid cell acts similar to an aquifer which can exchange liquid with the atmosphere above. If the bucket is sufficiently deep, with enough liquid in it, it can act as an ocean. The level of liquid $z$ within the bucket evolves according to surface fluxes as
\begin{equation}
	\frac{\dff z}{\dff t} = \rho_v \left(P-E\right) \rlap{,}
\end{equation}
where $\rho_v$ is the liquid density of the condensable component. Precipitation and condensation from the lowest atmospheric model layer contribute to $P$ whilst evaporation is denoted by $E$. The rate of evaporation from a bucket is described by the function
\begin{equation}\label{eq:bucket_model_modified}
	E = 
	\begin{cases} 
		E_0 & \qquad\textrm{when}\qquad z \geq z_\textrm{oc} \\ 
		\frac{z}{z_\textrm{oc}} E_0 & \qquad\textrm{when}\qquad z < z_\textrm{oc} \rlap{,}
	\end{cases}
\end{equation}
which denotes two evaporative regimes determined by the amount of moisture within the bucket. When the liquid depth in the bucket exceeds a critical value $z_\textrm{oc}$, which we denote the ocean threshold, evaporation is equivalent to that of an aquaplanet using a mixed-layer slab ocean, with the base rate of evaporation $E_0$ determined by 
\begin{equation}
	E_0 = c_p \rho C_D \left|\bs{v}_s\right|\left(q_s - q_\textrm{atm}\right) \rlap{.}
\end{equation}
The heat capacity at constant pressure $c_p$, heat transfer coefficient $C_D$ and horizontal surface wind magnitude $\left|\bs{v}_s\right|$ are used along with the specific humidity $q$ at the surface and lowest atmospheric level. Below the critical depth $z_\textrm{oc}$ evaporation is described by a linear ramp where $\lim_{z\to 0}E=0$. Each individual bucket is able to exchange moisture only with the overlying atmosphere and acts independently of one-another. Each bucket has an infinite capacity which means there is no overflowing and the total planetary moisture inventory is conserved. Unless otherwise stated $z_\textrm{oc}=375$~\si{\m} is used as a base configuration, derived from \citet{manabe1969-CLIMATE}.

Moist convection is represented using the quasi-equilibrium `simple Betts-Miller' (SBM) scheme \citep{frierson2007-Dynamics}. The SBM scheme relaxes the atmospheric relative humidity profile along a moist adiabat over a prescribed timescale \citep{betts1986-New, betts1986-Newa} when convective instability occurs, with the parametrisation using default values of $\textsc{rh}_\textsc{sbm}=70\%$ and $t_\textsc{sbm}=\;$\SI{7200}{\s}. The convection scheme provides one source of precipitation in the model, where warm, moist air is lifted until the parcel becomes saturated and condenses out. Clouds are not present in this configuration of \textit{Isca}, so any condensed moisture is immediately precipitated out. The convection scheme is coupled with a large-scale condensation scheme that removes excess vapour beyond saturation in each grid cell and permits the re-evaporation of falling precipitation according to the local humidity. \textit{Isca} provides two distinct precipitation outputs for convective and large-scale condensational rain, allowing their contribution to the local total precipitation to be separately attributed.
% See Stephen's note about  virtual temperature

Using the basic model configuration described above, we present a set of simulations where the initial liquid depth $z_0$ and ocean threshold $z_\textrm{oc}$ are adjusted. The initial liquid depth sets the planetary moisture inventory whilst the ocean threshold influences the ease with which the surface may dry as liquid evaporates from the bucket. The available values for these parameters are given below, providing 9 different simulations in which:
\begin{equation}\label{eq:ch4_initial_conditions}
	z_0 = \left\{
\begin{array}{r}
	1  \\
	10  \\
	100 
\end{array}
\right. \si{\m} 
\qquad\qquad
z_\textrm{oc} = \left\{
\begin{array}{r}
	100 \\
	200 \\
	375 
\end{array}
\right. \si{\m} 
\rlap{.}\nonumber
\end{equation}
The values of $z_0$ are less than the equivalent value for Earth at approximately \SI{2700}{\m} \citep{genda2016-Origin}, however are much closer to the currently predicted values for Titan, a tidally-locked body with respect to Saturn \citep{hayes2016-Lakes, cordier2021-Vertical}. Each simulation described above is run for a minimum of 2000 years (starting from year 0), allowing ample time for any large-scale effects to emerge. Since there is no seasonal cycle, short-term climatic variation is minimal thus data is output as annually-averaged values. The namelists used for running the models can be accessed at \texttt{\url{https://github.com/daw538/williams2025-hydrocycle}}.

\section{Results}
We focus on two simulations, those using $z_0=10$ and \SI{100}{\m} initial depth and ocean threshold $z_\textrm{oc}=375$\si{\m} (as per \citet{manabe1969-CLIMATE}). These two simulations are chosen since their long-term evolutions are very different, representing two distinct climatological evolutions. The simulation initialised with $z_0=$\SI{10}{\m} liquid depth is denoted the `drier' simulation and that with $z_0=$\SI{100}{\m} the `wetter' simulation.
\begin{figure*}[htbp]
	\centering
	\includegraphics[width=0.8\textwidth]{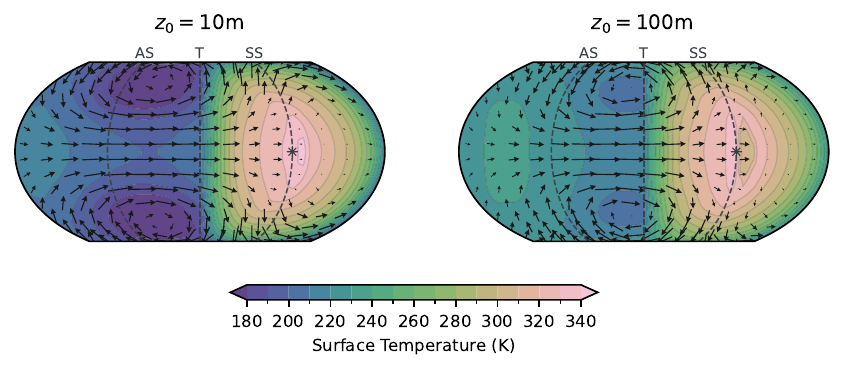}
	\caption[Surface temperature and \SI{500}{\hecto\Pa} wind field after 2000 years]{Surface temperature distribution after 2000 years, given in filled contours. Wind field at \SI{500}{\hecto\Pa} is given by black quivers. Initial liquid depth in the bucket model is given by $z_0$. The substellar point is marked with an asterisk, while dashed grey meridians represent key longitudes: the substellar longitude (SS), western terminator (T) and antistellar longitude.}
	\label{fig:4-2k_temperature}
\end{figure*}

\subsection{Long-term Global Climatology}

\begin{figure*}[htbp]
	\centering
	\includegraphics[width=0.8\textwidth]{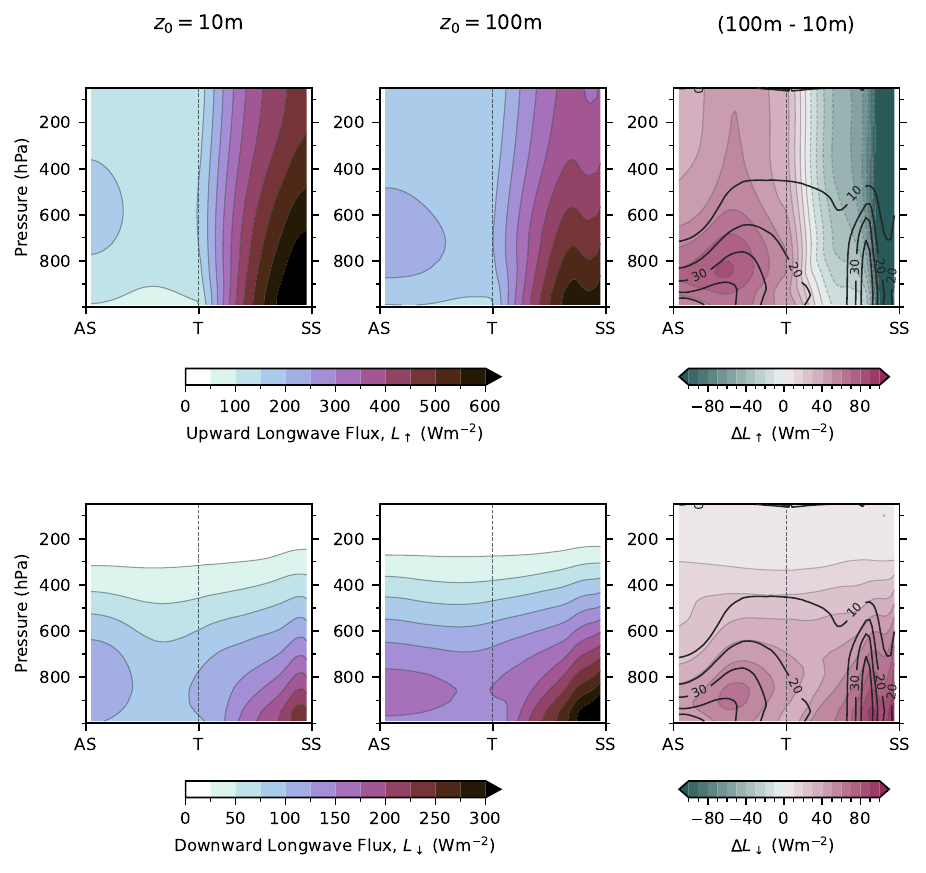}
	\caption[Contribution by water to long-wave heating in the night side hemisphere]{A comparison of the upwards and downwards long-wave radiative fluxes (left and middle column) after 2000 years using a tidally-locked co-ordinate system. AS, T and SS denote the antistellar point, terminator and substellar point respectively. The right-hand column shows the difference in fluxes in filled contours, which are overlain  by solid black contours that indicate the percentage increase of specific humidity in the \SI{100}{\m} versus the \SI{10}{\m} simulation.}
	\label{fig:4-water_radiative_heating}
\end{figure*}

After two millennia both simulations display a classically tidally-locked surface temperature structure, shown in Figure~\ref{fig:4-2k_temperature} where the day side is much hotter than the night side; lacking the longitudinal symmetry that would be present on a planet without tidal-locking such as Earth. The drier simulation shows a peak surface temperature of \SI{343}{\K} downstream (east) of the substellar point, with temperature then decreasing monotonically towards the terminator. In the wetter simulation, a local \textit{minimum} of \SI{304}{\K} exists at the same location, while even larger differences between the two simulations can be found in the night-time hemisphere. The surface cold spots for both simulations are found in the high-latitude regions upstream (west) or the terminator, being \SI{165}{\K} and \SI{205}{\K} for the drier and wetter simulations respectively. The surface cold-spots in each simulation are associated with cyclonic gyres in the \SI{500}{\hPa} wind field, with a strong equatorial jet separating these two regions west of the terminator that helps advect heat around the planet. The surface temperature and mid-tropospheric wind structure arises from the influence of planetary waves \citep{yang2013-STABILIZING}, with a wavenumber-1 stationary wave pattern characteristic of tidally-locked planets with a 10-day rotation period \citep{hammond2018-Wavemean}. 

The smaller range of surface temperatures seen on the wetter planet (\SI{131}{\K} versus \SI{179}{\K} on the drier planet) is attributed to an enhanced night-side greenhouse effect. With more atmospheric moisture present, day-night heat transport is enhanced as a greater amount of long-wave radiation is absorbed and re-emitted towards the surface, as shown in Figure~\ref{fig:4-water_radiative_heating}. The impact of latent heating on the heat transport is small between the two hemispheres. These results agree with literature and demonstrate that dry tidally-locked planets are expected to have higher temperature gradients than their aquaplanet counterparts; with our simulation reflecting intermediate states.
\begin{figure*}[htbp]
	\centering
	\includegraphics[width=0.8\textwidth]{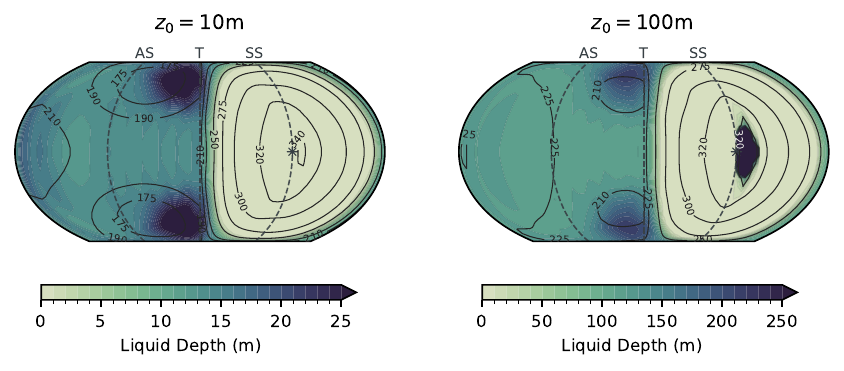}
	\caption[Surface hydrology after 2000 years]{Distribution of surface hydrology after 2000 years. Filled contours represent liquid depth present within the surface buckets whilst surface temperature is depicted with labelled solid contours. }
	\label{fig:4-2k_liquid_depth}
	\includegraphics[width=0.8\textwidth]{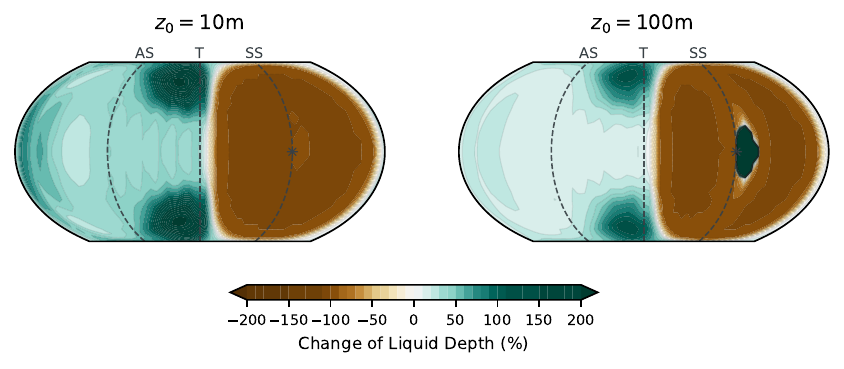}
	\caption[Relative change in surface hydrology after 2000 years]{Relative change of liquid present in surface buckets with respect to initialised depth. Contours represent percentage change after 2000 years.}
	\label{fig:4-2k_liquid_depth_change}
\end{figure*}

The surface hydrology after 2000 years is shown by Figures~\ref{fig:4-2k_liquid_depth}~and~\ref{fig:4-2k_liquid_depth_change}, which show the first indications of significant deviation between the two simulations. In common is the development of an `eyeball' climatology consistent with expectations for a moist tidally-locked planet \citep{wordsworth2022-Atmospheres}. Liquid has evaporated from the surface of the permanently illuminated day side and transported by the atmospheric circulation to the night side, where it precipitates; that is to say, the condensable is cold-trapped on the night side. This explains the deeper regions of liquid near the high-latitude cold-spots since vapour easily condenses out from the cold atmosphere above. The key difference between the two simulations lies just east of the substellar point: after 2000 years the wetter simulation has developed an isolated deep body of water, which we denote the `substellar lake'. The remaining day side is otherwise barren, with the similarities and differences between the two simulations more clearly presented by the relative changes in surface liquid depth since initialisation in Figure~\ref{fig:4-2k_liquid_depth_change}. The enhanced precipitation just east of the substellar point in the wetter case is similar to that occurring in the Madden--Julian oscillation on Earth, for there too precipitation forms just east of the heating \citep{Adames_Wallace14, Vallis21} causing the disturbance to propagate eastward, although in the tidally-locked case the heating is strongly tied to the substellar point preventing propagation.

\begin{figure*}[htbp]
	\centering\includegraphics[width=0.8\textwidth]{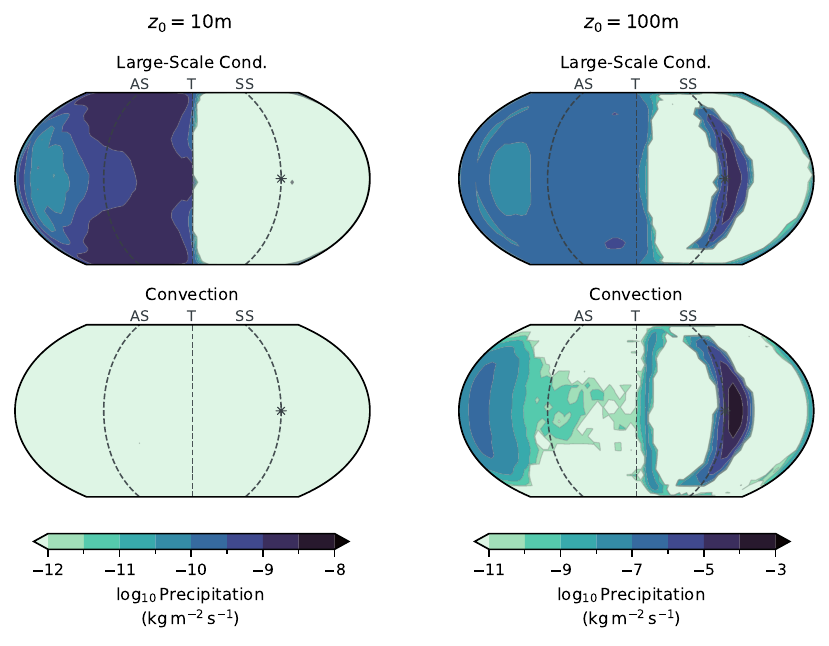}
	\caption[Precipitation patterns after 2000 years]{Comparison of precipitation distribution after 2000 years including breakdown of precipitation caused by the large-scale condensation or convection schemes. Contours depict precipitation rate on a logarithmic scale.}
	\label{fig:4-2k_precipitation}
\end{figure*}

In the absence of horizontal surface transport, the only mechanisms that allow liquid depth to increase in the simulations are precipitation and surface condensation. With surface temperatures near the substellar point being so high, formation via surface condensation is considered very unlikely, leaving precipitation as the sole source of moisture that can replenish the substellar lake. The model configuration in \textit{Isca} permits precipitation from either large-scale condensation or the convection scheme, so we can assess the relative contribution by each source in both simulations, shown in Figure~\ref{fig:4-2k_precipitation}. 

The quantity of total precipitation is orders of magnitude larger in the wetter simulation due to the greater surface reservoir of moisture, with peak total precipitation rates of \SI{2e-8}{\kg\per\m\squared\per\s} (drier) and \SI{2e-4}{\kg\per\m\squared\per\s} (wetter). Precipitation in the drier simulation is limited to the night-time hemisphere, which is attributed to large-scale condensation resulting from the advection of moist air to the cooler night side of the planet. In the wetter simulation, precipitation is attributed to both large-scale and convective sources. On the night side, large-scale precipitation provides the substantial contribution via the cold-trapping mechanism. Additionally however is an elongated region in the substellar region in the vicinity of the lake, which is also seen as a strong contribution from convective precipitation---it is notable that the highest intensities anywhere on the planet for either source of precipitation are over the substellar lake region. The eastward shift of the peak relative to the substellar point results from the equatorial wave dynamics that produce a convergent region of comparatively moisture-laden air in the troposphere \hbox{\citep{yang2013-STABILIZING}}, a process that may be similar to that occurring in Earth's Madden--Julian oscillation \citet{Vallis_Penn20, Vallis21}. The convergence and associated upward motion allows saturation to be satisfied, explaining the contribution from both large-scale and convective sources. Convective precipitation is also found to a lesser degree near the terminator. With the exception of the substellar lake, results presented thus far are fairly comparable with existing literature on tidally-locked planet. The substellar lake however is an unusual feature, and further investigation of its evolution and persistence follow.

\begin{figure*}[htbp]
	\centering
	\includegraphics[width=0.8\textwidth]{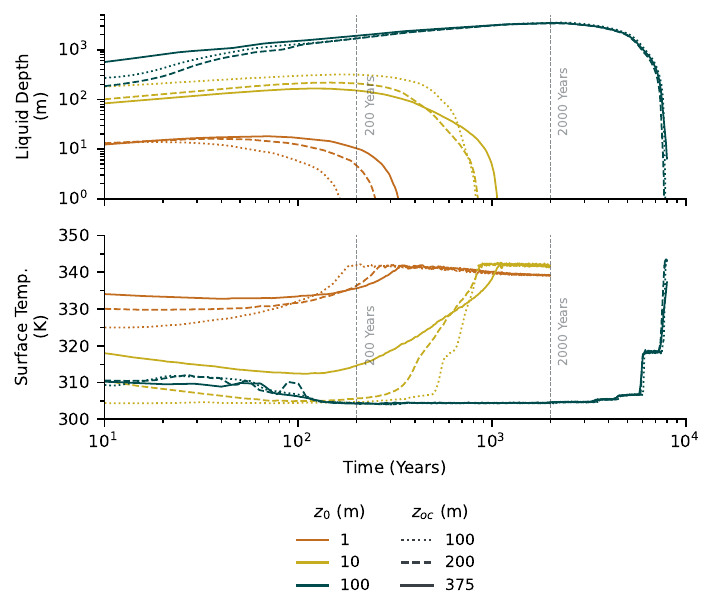}
	\caption[Evolution of lake region liquid depth and surface temperature]{Evolution of surface quantities in the lake region defined by \SI{5}{\degree} to \SI{15}{\degree} longitude and \SI{-5}{\degree} to \SI{5}{\degree} latitude over the first 2000 years, presented as a 10-year rolling mean for each of the 9 model permutations. Initial liquid depth $z_0$ is indicated by line colour whilst ocean threshold $z_\textrm{oc}$ is represented by the line style.}
	\label{fig:4-lake_region_evolution}
\end{figure*}

\subsection{Evolution of the Substellar Region}
Our analysis so far indicates a clear distinction in long-term evolution between tidally-locked planets initialised with a lower ($z_0=10$~\si{\m}) and higher ($z_0=100$~\si{\m}) moisture inventory. Given both planets are initialised with a finite global ocean, of great interest is how they diverge and evolve over the subsequent two millennia. From Figure~\ref{fig:4-2k_liquid_depth_change} one can deduce that the substellar lake develops at the expense of the surrounding daytime surface region. For the drier simulation, the evolution is less clear---did the entire hemisphere dry uniformly or did a transient substellar lake form before drying out in turn after the regions surrounding it?

Figure~\ref{fig:4-lake_region_evolution} presents the surface evolution of the lake region over multiple millennia, with quantities averaged over the deepest portion of the lake: \SI{5}{\degree} to \SI{15}{\degree} longitude and \SI{-5}{\degree} to \SI{5}{\degree} latitude, where \SI{0}{\degree} lat-lon is defined by the substellar point . All nine simulation configurations, varying initial surface moisture budget and ocean threshold depth, are presented. The initial liquid depth provides the strongest effect on surface evolution. In all cases the depth of liquid in the substellar lake region initially increases before the simulations begin to diverge. For the drier set of simulations ($z_0=1$ and \SI{10}{\m}) the deepest value is attained between a couple of decades to a couple of centuries after initialisation. After this drying accelerates until the lake fully evaporates, with total lake lifespan varying from 150--350 years ($z_0=$\;\SI{1}{\m}) to 1000--1300 years ($z_0=$\;\SI{10}{\m}). The timescales involved for the wetter simulation ($z_0=$\;\SI{100}{\m}) are substantially different however, with the lake continuing to deepen even after 2000 years. Extended simulation runs indicate a lifespan of approximately 8000 years for these simulations, making the substellar lake a persistent but nonetheless transient feature. The effect of an adjusted ocean threshold $z_\textrm{oc}$ in the bucket model is to modify the rate at which the substellar lake region initially deepens and subsequently evaporates away. When $z_\textrm{oc}$ is lower, a higher rate of evaporation can occur when the bucket, as per Equation~\ref{eq:bucket_model_modified} and resulting in an increased rate of drying. The signal is clearest when the initial liquid depth is small: halving the value of $z_\textrm{oc}$ shortens the lifetime of the lake in the $z_0=1$\si{\m} simulations by 100 years. When the initial depth is larger, increased evaporation in the daytime region encircling the substellar point (closer to the terminator) may in fact accelerate the accumulation of liquid in the substellar lake, leading to the more mixed behaviour during the lake growth period when $z_0=$\;\SI{10}{\m} and $z_0=$\;\SI{100}{\m}.

The influence of a substellar lake on the local climate is visible in the impact on the surface temperature via precipitation and evaporative cooling (latent heat exchange). The cooling effect of the lake is as large as \SI{30}{\K} relative to the surrounding environs, seen as a local surface temperature minimum in Figure~\ref{fig:4-2k_temperature}. As the lake shrinks and depth decreases, so does its capacity to moderate the local surface conditions, explaining the anti-correlation between lake depth and surface temperature. All simulations result in a surface maximum temperature in excess \SI{340}{\K} once the substellar lake has disappeared. By assessing the temporal evolution of liquid depth and surface temperature in the substellar lake region, Figure~\ref{fig:4-lake_region_evolution} also addresses the earlier question about whether the lake is a permanent or transient feature. Evidently it is transient, with lifespan being controlled by initial surface depth and to a lesser extent, the ocean threshold. The relation between initial depth and lifespan is also non-linear.

\begin{figure*}[htbp]
	\centering
	\includegraphics[width=0.8\textwidth]{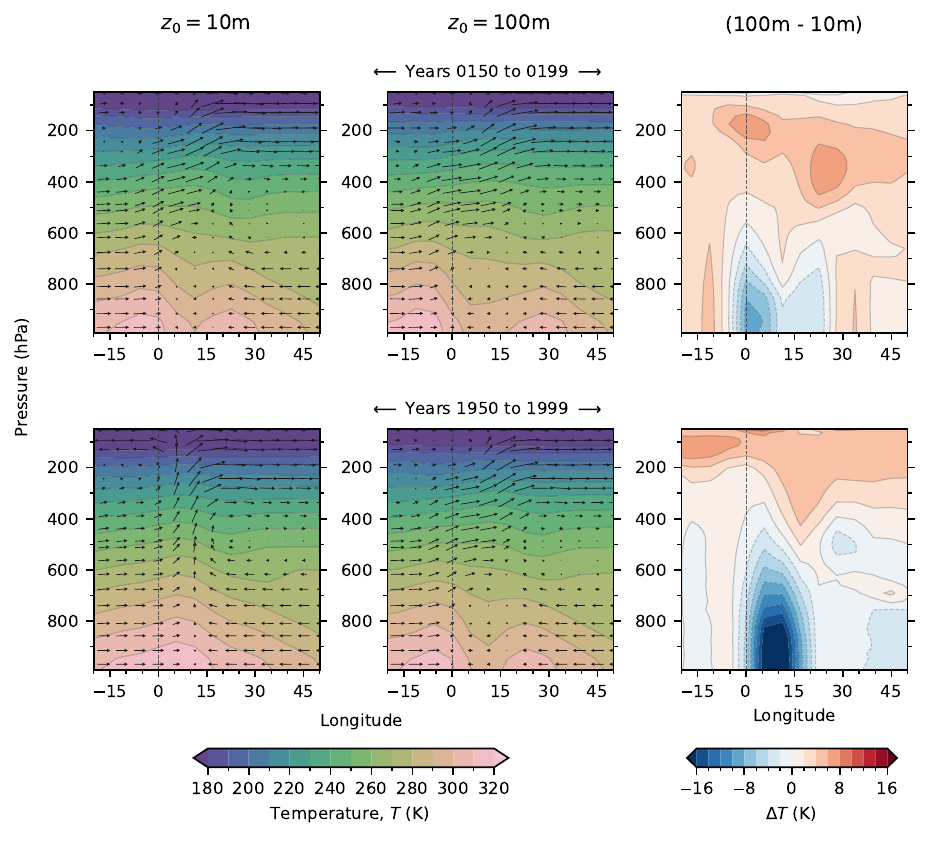}
	\caption[Lake region equatorial cross-section of temperature]{Atmospheric cross-section of temperature along the equator between \SI{-20}{\degree} and \SI{+50}{\degree} longitude relative to the substellar point. Left and centre columns represent the simulations with initialised liquid depth of \SI{10}{\m} and \SI{100}{\m}, where temperature is represented in filled coloured contours and zonal wind by black arrows. Right-hand column depicts the difference between the two simulations, as shown by filled blue-red contours. Rows represent two climatologies: early evolution (top) between 150-199 years; long-term state (bottom) between years 1950 and 1999.}
	\label{fig:4-lake_region_temp}
\end{figure*}

Knowing that there is a strong non-linear relationship between initial depth and substellar lake lifespan, is it possible to elucidate from the early climatic evolution whether the lake will dry over a short timescale or be quasi-stable over thousands of years? Figure~\ref{fig:4-lake_region_temp} presents the state of the troposphere above the substellar lake in early (years 150--200) and later-stage (years 1950--2000) evolution as a means of discerning the mechanism that sustains the lake over longer periods. During early evolution the drier ($z_0=$\SI{10}{\m}) and wetter ($z_0=$\SI{100}{\m}) simulations show similar vertical atmospheric structures in both temperature and meridional wind profile, with the largest difference in the highest surface temperature near the substellar point. An equatorial westerly flow rises from \SI{600}{\hecto\Pa} to \SI{200}{\hecto\Pa} over a less coherent convergent region above the lake between \SI{0}{\degree} and \SI{20}{\degree} longitude. The lower level temperature profile is modified by evaporative cooling over the lake whilst higher up the temperature structure is more stratified. After 2000 years, differences have appeared in the vertical structure, with the wetter simulation remaining somewhat unchanged but the drier simulation experiencing a change both to the temperature and wind structure over the lake. Disappearance of the substellar lake has removed the local temperature minimum at the surface (as seen in Figures~\ref{fig:4-2k_temperature} and~\ref{fig:4-2k_liquid_depth}), with temperatures now \SI{20}{\K} higher than the wetter simulation over the former lake region. Convergence east of the substellar point has also become more coherent with a stronger updraught present than before, whilst the upper troposphere has cooled marginally as a result of reduced latent heating from moisture lifted near the surface. 

\begin{figure*}[htbp]
	\centering
	\includegraphics[width=0.8\textwidth]{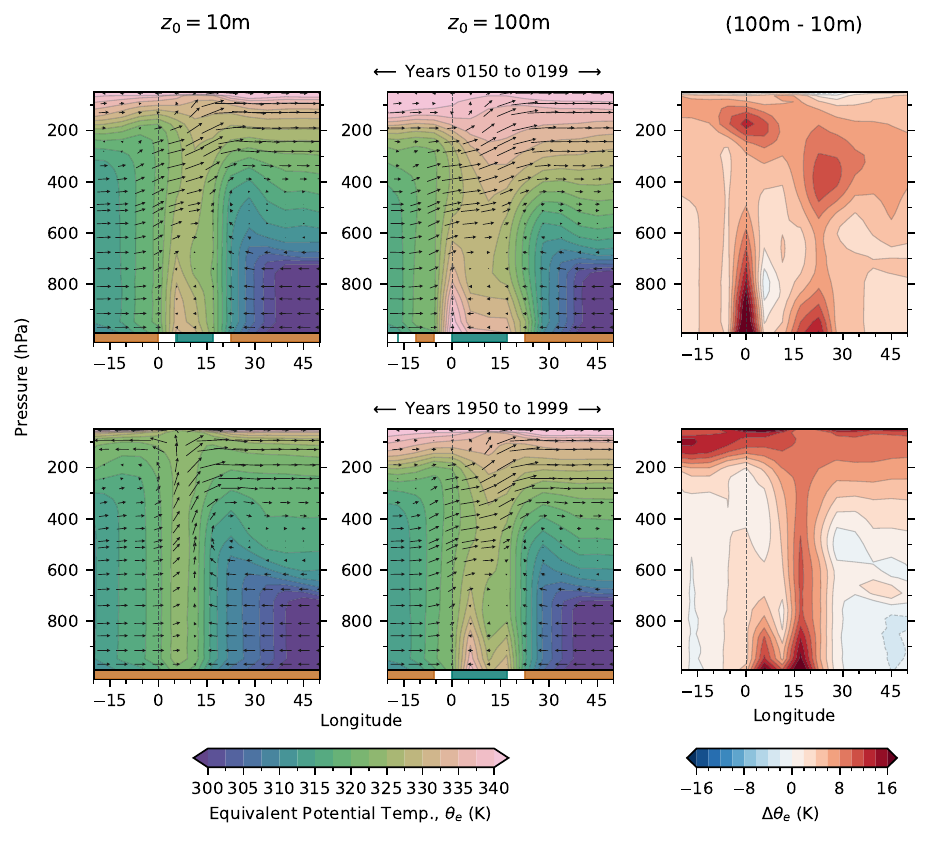}
	\caption[Lake region equatorial cross-section of equivalent potential temperature]{Atmospheric cross-section of equivalent potential temperature along the equator between \SI{-20}{\degree} and \SI{+50}{\degree} longitude relative to the substellar point. Left and centre columns represent the simulations with initialised liquid depth of \SI{10}{\m} \SI{100}{\m}, where temperature is represented in filled coloured contours and zonal wind by black arrows. Coloured bars at surface depict if convective inhibition in simulation is a positive (teal) or negative (orange) value. Right-hand column depicts the difference between the two simulations, as shown by filled blue-red contours. Rows represent two climatologies: early evolution (top) between 150-199 years; long-term state (bottom) between years 1950 and 1999.}
	\label{fig:4-lake_region_thetae}
\end{figure*}

Alternatively, one can assess the equivalent potential temperature $\theta_e$, which includes contribution from gravitational potential and latent head, to understand differences between the evolution of each simulation---presented in Figure~\ref{fig:4-lake_region_thetae}. Since the definition includes a moisture component, we anticipate a higher average $\theta_e$ in the wetter simulation, so we look for more subtle differences. Both simulations display a stably stratified region either side of the substellar lake region during early evolution. The structure above the lake is substantially different, with similar values of $\theta_e$ at the surface and in the upper troposphere; the resulting negative gradient indicating a column of air highly unstable to convection. While both simulations show this distribution, the differences are more pronounced than in Figure~\ref{fig:4-lake_region_temp}, which a much wider `neck' of instability indicating a larger region prone to convection. After 2000 years the drier simulation has lost the inversion of equivalent potential temperature; instead showing a neutrally-stratified column over the former lake region and agreeing with past results in \citet{pierrehumbert2019-Atmospheric}. The negative gradient and neck of unstable air remains the wetter simulation, indicative of a stronger convective circulation that is likely helping to prolong the substellar lake lifetime through production of precipitation.

To assess the extent of the convective region, one can use convective inhibition (CIN) to diagnose if conditions at the surface are conducive to strong convection. CIN is measure of the energy required to lift a parcel of air from the surface to the lifting condensation level and thence the level of free convection (LFC), where the parcel will then ascend without further energy input. It is defined using 
\begin{equation}
	\textrm{CIN} = -\int_{p_0}^{p_\textrm{LFC}} R_d \left(T_{v,p} - T_{v,e}\right)\dff\ln p \rlap{,}
\end{equation}
where $R_d$ is the dry gas constant and $T_v$ is the virtual temperature (additional subscripts $T_{v,p}$ and $T_{v,e}$ consider the parcel and environmental values), given in turn (in \textit{Isca}) as
\begin{equation}
	T_v = T \left(1+q\left(\frac{R_v}{R_d}-1\right)\right) \rlap{,}
\end{equation}
where $q$ is the specific humidity and $R_v$ is the moist gas constant. Negative CIN represents an energy cap that must be overcome for a parcel to convect freely. A positive value is equivalent to no cap, indicating convection occurs freely from the surface and spontaneously. 

Significant non-zero values of calculated CIN are shown at the surface in Figure~\ref{fig:4-lake_region_thetae}. Areas of positive CIN underlie the inversions of equivalent potential temperature, indicating a highly unstable environment over the substellar lake and in contrast to the substantial cap existing over the wider region. The wetter simulation sustains a highly unstable surface layer that extends beyond the spatial extent of the substellar lake, suggesting an active convective plume over a wider region than the lake. The region of positive CIN in the drier simulation is narrower, even during early evolution, and eventually disappears. We deduce from analysis of CIN and $\theta_e$ that the wetter simulation sustains a much larger and more powerful convective region near the substellar lake; sufficient to pull greater amounts of moisture into the convective plume that subsequently condenses and precipitates out as it is lifted aloft.

\begin{figure*}[htbp]
	\centering
	\includegraphics[width=0.8\textwidth]{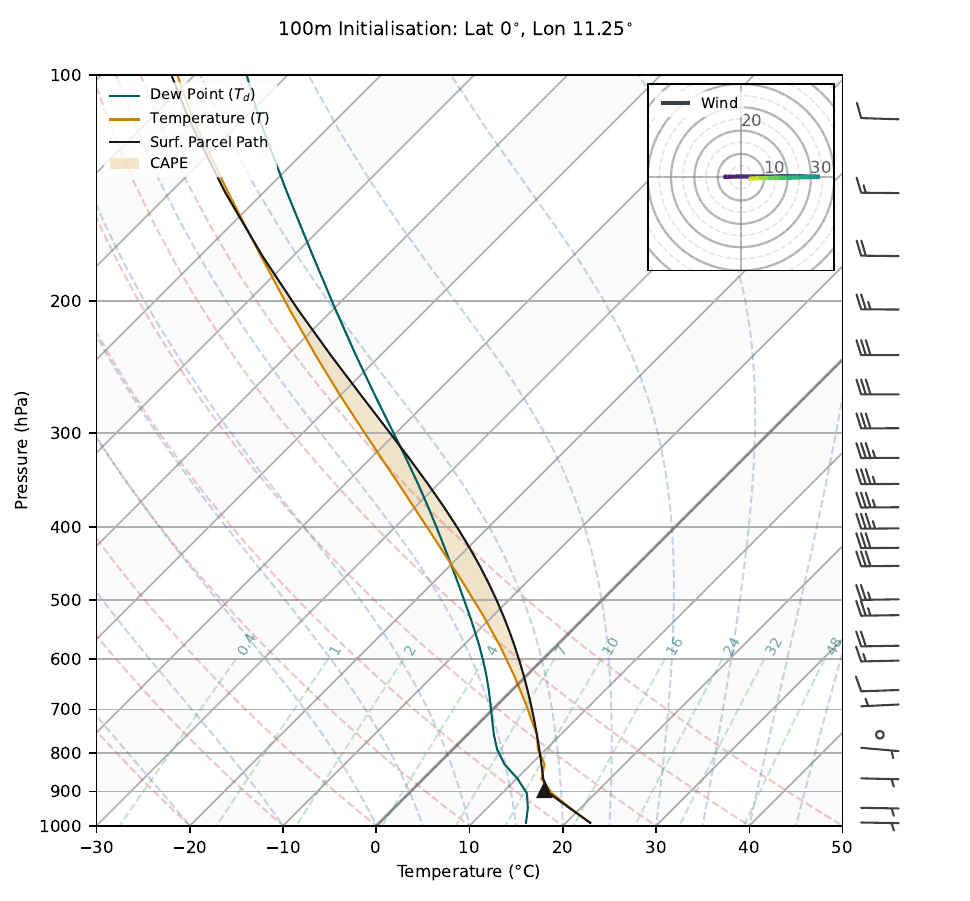}
	\caption[Lake region skew-T diagram for \SI{100}{\m} initial surface hydrology]{Skew-T diagram of the simulation initialised with \SI{100}{\m} surface hydrology after 2000 years, specified near the centre of the lake region at \SI{11.25}{\degree}. Orange and teal solid lines are the temperature and dew point vertical profiles, with the black line representing the path taken by a parcel at the surface if lifted adiabatically. Black triangle depicts the lifting condensation level, with shaded orange and teal regions respectively depicting convective available potential energy and convective inhibition. Faint background lines represent atmospheric quantities: pressure (grey horizontal); temperature (grey diagonal); dry adiabats (dashed blue); moist adiabats (dashed red); isohumes, lines of constant mixing ratio in \si{\g\per\kg} (dashed green). Barbs on right show wind direction and strength, with each full barb representing \SI{5}{\m\per\s}. A hodograph inset shows wind direction with height: colours become lighter at higher altitudes. }
	\label{fig:4-tephigram_wet}
\end{figure*}

\begin{figure*}[htbp]
	\centering
	\includegraphics[width=0.8\textwidth]{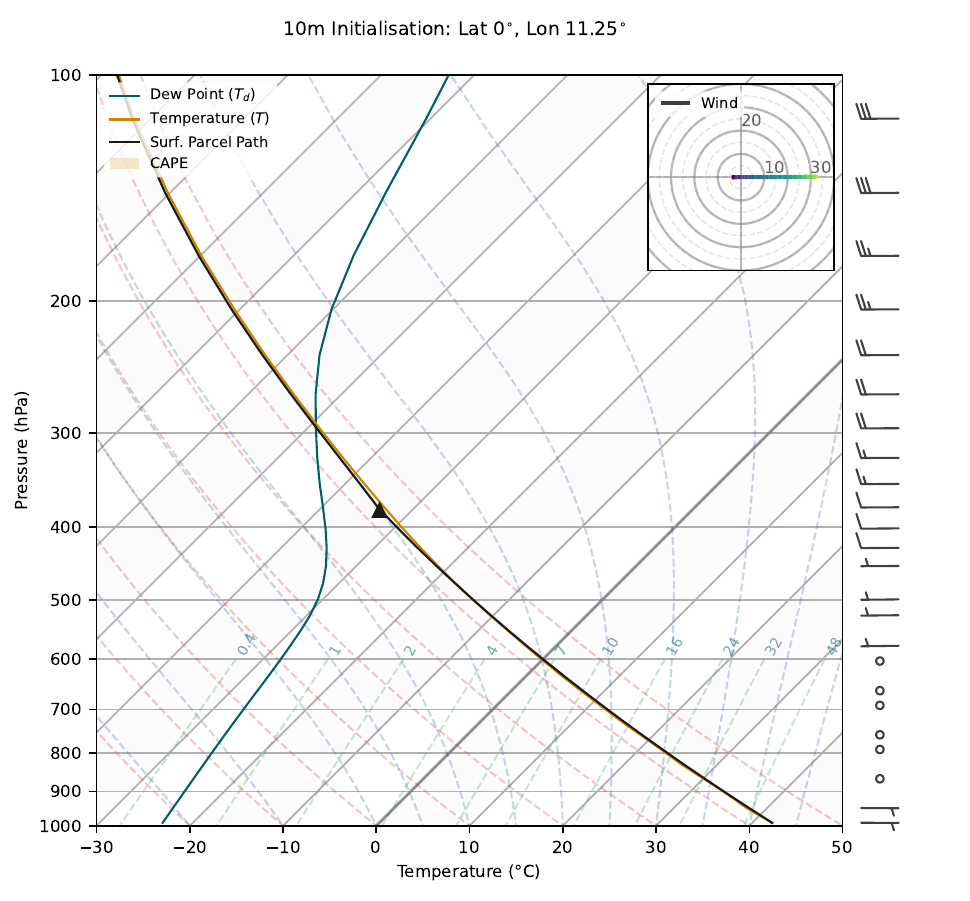}
    \caption{As for Fig.\ \ref{fig:4-tephigram_wet} but for an initial \SI{10}{\m} depth initial surface hydrology.}
	\label{fig:4-tephigram_dry}
\end{figure*}

We make further assessment of atmospheric stability in the substellar lake region through use of thermodynamic diagrams such as a {skew-T} plot (or tephigram). Typically used for analysing radiosonde data on Earth, they are an useful tool for understanding the behaviour of an ascending air parcel. {Skew-T} diagrams after 2000 years are constructed with (\SI{0}{\degree} N, \SI{11.25}{\degree} E) as the reference lake location, shown for each simulation in Figures~\ref{fig:4-tephigram_wet} and~\ref{fig:4-tephigram_dry}. The two figures are considerably different and we consider a rising air parcel from each in turn. In the wetter simulation (Figures~\ref{fig:4-tephigram_wet}) a parcel of air lifted from the lake surface at \SI{23}{\degreeCelsius} (\SI{296}{\K}) and dew point of \SI{16}{\degreeCelsius} (\SI{289}{\K}) can be lifted adiabatically by convection to a LCL at approximately \SI{900}{\hPa}. If the parcel ascends further, moisture will be precipitated out and returned to the lake directly below, providing a mechanism for the lake to replenish itself. Once the lake is sufficiently deep, the requirement to sequester moisture from convergent air at the substellar point reduces and the lake can (to an extent) sustain itself. The dynamics of the lake and overlying lower troposphere may be considered a local phenomenon that explains the high contribution from convection to the total precipitation in Figure~\ref{fig:4-2k_precipitation}.

By contrast, the drier simulation presents a very different profile (Figure~\ref{fig:4-tephigram_dry}). A surface air parcel from the former lake region at over \SI{42}{\degreeCelsius} (\SI{315}{\K}) and dew temperature of \SI{-23}{\degreeCelsius} (\SI{250}{\K})---the combination indicating a very arid environment---must be lifted to approximately \SI{375}{\hPa} in order to reach the LCL. Any moisture lifted aloft or introduced by surface convergence will be released into the large-scale westerly flow (Figure~\ref{fig:4-lake_region_temp}) before condensing, providing a conveyor for moisture to be transported and lost from the substellar region before precipitation occurs. This mechanism reflects the first half of the cold-trapping cycle and indicates that any moisture that escapes from the night side cold-trap will invariably be returned there. In short, once the substellar lake is lost it cannot be reformed.

\subsection{Lifetime of the Substellar Lake}
One question that inevitably arises is that of the prolonged (but finite) longevity of the substellar lake in wetter climates. Given that planets with a drier hydrology ($z_0=1$ or \SI{10}{\m}) are unable to retain unable to retain a substellar lake beyond a few hundred years, is there a depth beyond which the lake transitions from being short-lived ($\sim$100 years) to lasting many millennia before drying out, or even becoming permanent? We analyse the surface conditions in the vicinity of the lake to probe these questions.

\begin{figure*}[htbp]
	\centering
	\includegraphics[width=0.8\textwidth]{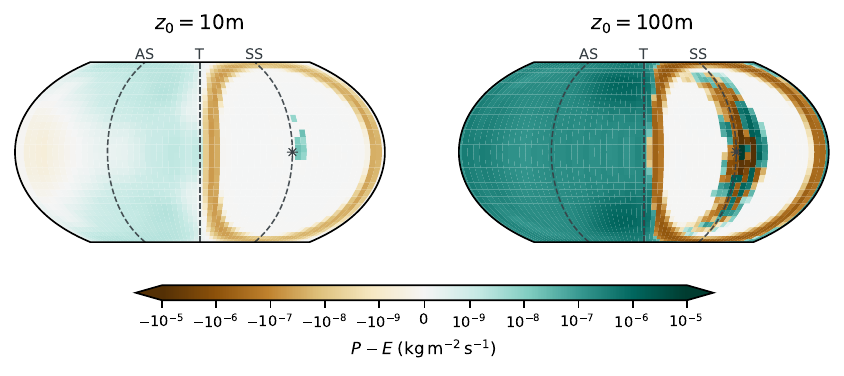}
	\caption[Surface moisture deficit ($P-E$) after 2000 years]{Comparison of surface moisture deficit after 2000 years, given by total precipitation minus evaporation ($P-E$). Shading is teal where precipitation exceeds evaporation rate and brown when evaporation exceeds precipitation. Colourbar uses a symmetric log scale to emphasize small values and deviations around zero. Between $-10^{-9}$ and $+10^{-9}$ the scale is linear whilst outside of this range it is logarithmic in both the positive and negative.}
	\label{fig:4-2k_p_minus_e}
\end{figure*}

Evolution of the liquid depth in the surface bucket hydrology is ultimately driven by the surface moisture deficit (i.e., the difference between precipitation and evaporation ($P-E$)), which is shown after 2000 years in Figure~\ref{fig:4-2k_p_minus_e} for the drier and wetter simulations---the former lost its lake over a millennium before whilst the latter still retains one. Positive $P-E$ indicate the mass of liquid in the surface buckets is increasing---\textit{i.e.}~the lake is becoming deeper---whilst negative indicates the opposite. Averaging over the previously defined lake region provides $P-E=$ \SI{4.842e-6}{\kg\per\m\squared\per\s} for the substellar region, or equivalently the depth of the lake in the wetter simulation after 2000 years is increasing by \SI{0.15}{\m} per year. The moisture around the lake is mostly supplied by a converging surface flow originating near the terminator. The equivalent figure averaged over the night side is \SI{2.865e-7}{\kg\per\m\squared\per\s} (\SI{0.09}{\m} per year). The rate of increase is very slow, but it alone is not enough to indicate if the lake is a permanent feature or not.

\begin{figure*}[htbp]
	\centering
	\includegraphics[width=0.8\textwidth]{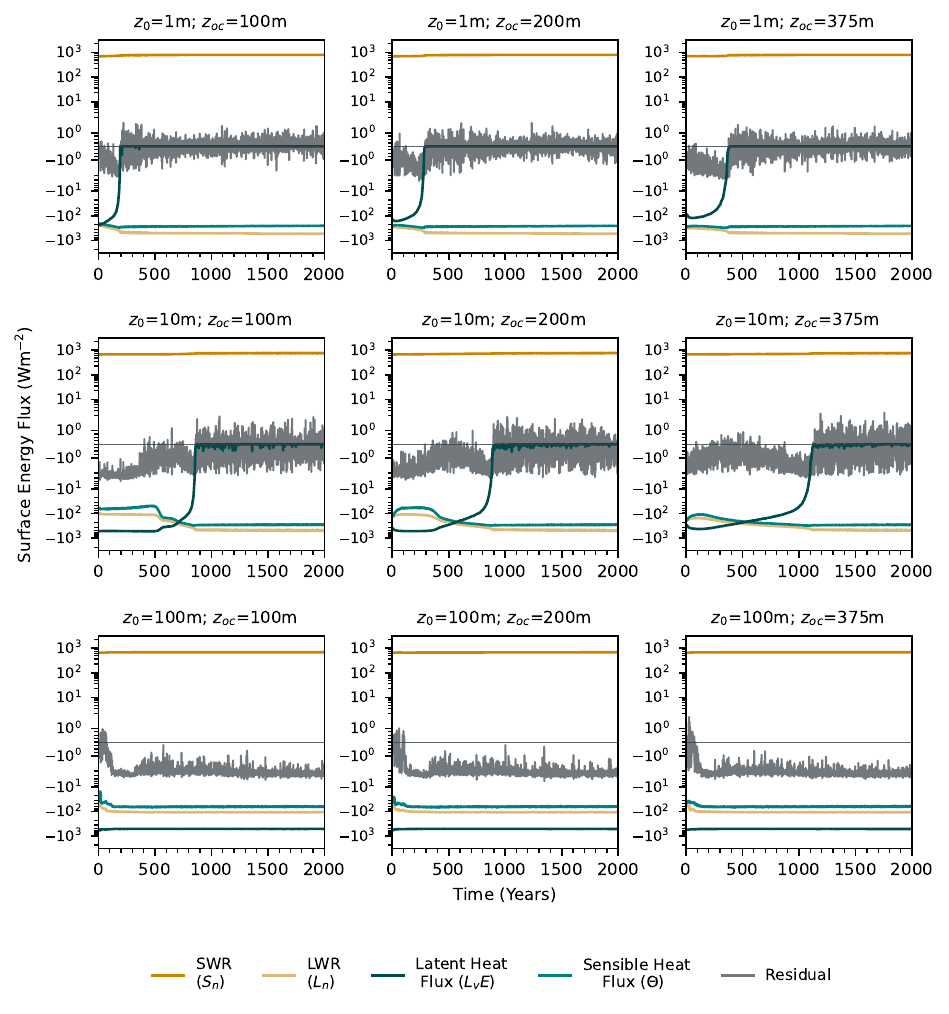}
	\caption[Evolution of substellar lake region surface fluxes]{Surface energy fluxes averaged over the substellar lake region. Fluxes are presented on a symmetric log scale to emphasize small values and deviations around zero. Between $-10^{0}$ and $+10^{0}$ the scale is linear whilst outside of this range it is logarithmic in both the positive and negative directions. Net short-wave and long-wave radiation fluxes are shown in light and dark orange, whilst the sensible and latent heat fluxes are shown in light and dark teal respectively. The residual is shown in grey.}
	\label{fig:4-evo_lake_surf_energy_balance}
\end{figure*}

Figure~\ref{fig:4-lake_region_evolution} helped establish that once the substellar lake is lost by evaporation it does not reappear. Any ingression of moisture back onto the day side is likely to be returned by the cold-trapping mechanism identified previously with Figure~\ref{fig:4-tephigram_dry}. If the expected ultimate climatology is an `eyeball' planet with moisture trapped on the night side of the planet, the local surface energy balance over the substellar lake will inform us if an equilibrium state has been achieved. Surface energy balance is calculated thus:
\begin{equation}
	\frac{\dff Q}{\dff t} = R_n - G - \Theta - L_v E \rlap{,}
\end{equation}
where Q is the total heat within the surface layer. Terms on the RHS are directionally defined as net surface irradiance $R_n$ (down), ground heat flux $G$ (up), sensible heat flux $\theta$ (up), and latent heat flux $L_V E$ (up). Irradiance can additionally be split into long-wave ($L_n$) and short-wave ($S_n$) components. While the impact of precipitation is ignored, a non-zero surface energy balance will still help indicate if the lake is a quasi-stable (transient) feature. Figure~\ref{fig:4-evo_lake_surf_energy_balance} provides an assessment of the surface energy balance of the substellar lake region for each of the nine simulations, with the radiation $R$ split into net short ($S_n$) and long-wave ($L_n$) components along with sensible ($\Theta$) and latent ($L_v E$) heat fluxes.

In the driest simulation ($z_0=1$~\si{\m}) an equilibrium state is achieved after a few hundred years, with a lower ocean threshold increasing the rate of convergence to an equilibrium. While the residual fluctuates it averages to zero over climatological timescales. As expected, the dry substellar region has no contribution from latent heat exchange and the energy balance is dominated by sensible heat flux. For an intermediate hydrology of $z_0=10$\si{\m}, the development of a substellar lake allows latent heat exchange to exceed sensible heat by an order of magnitude. Whilst a low ocean threshold allows equilibrium to be achieved sooner, it lengthens the duration over which latent heat fluxes dominate, likely because the rate of evaporation is also increased and the surface approximates the evaporative flux is enhanced at lower liquid depths (see Equation~\ref{eq:bucket_model_modified}) when $z_\textrm{oc}$ is reduced. Once the lake dries out latent heat fluxes become negligible and the residual again fluctuates around zero. For the wetter simulation ($z_0=100$\si{\m}) evolution is different: as expected latent heat fluxes dominate, yet after the initial spin-up period (100--200 years) there are no significant changes to any of the surface fluxes with $z_\textrm{oc}$ having no impact over the comparison period. After 2000 years the residual for all three variants of the wetter simulation remains non-negligible at approximately \SI{-2}{\radflux}. Whilst small, it indicates that the substellar region is not in a true steady-state equilibrium. This confirms the earlier findings from the extended simulation runs in Figure~\ref{fig:4-lake_region_evolution} that the lake is not a permanent feature, but rather one with a very long lifespan of thousands of years. Despite the substellar lake being replenished by precipitation due to the low LCL and local convective system, a very small proportion of moisture escapes into the large-scale planetary jet and is cold-trapped on the night side. Over thousands of years this slowly depletes the lake of source material---any moisture subsequently returned to the substellar region will be removed before it can form a lake by the resulting local conditions and large-scale circulation.

\section{Summary and Discussion}
Using the flexible climate modelling framework \textit{Isca}, we have explored how the climate of moist tidally-locked terrestrial planets evolve when ground hydrology with a finite moisture inventory is included. A summary of the key findings follows:

\begin{enumerate}[itemsep=0ex]
	\item The long-term evolution of climatology on a tidally-locked terrestrial planet can be strongly impacted by the initial surface moisture inventory. The final equilibrium state is a planet with `eyeball' climatology---a barren day side and wet night side---although the timescales to reach this state vary by orders of magnitude dependent upon the initial state.
	\item Before transition to the eyeball climatology, a lake can form just east of the substellar point. The lifetime of the lake depends non-linearly on the hydrological inventory. Planets with a shallower initial global layer of liquid ($z_0\lesssim10$\si{\m}) develop and lose the lake within a couple of hundred years. Planets with a deeper initial layer ($z_0\gtrsim10$\si{\m}) can retain the lake for many thousands of years. The presence of a continent at the substellar point might however prevent the formation of any such lake \citep{lewis2018-Influence}.
	\item The presence of a substellar lake modifies the local surface conditions and the atmosphere above it. Surface temperatures are moderated down by approximately \SI{30}{\K} due to evaporative cooling, with the associated moist convection and enhanced day-night heat transport increasing temperatures higher in the atmospheric column.
	\item The lifetime of the substellar lake is prolonged by a convectively unstable surface region and low lifting condensation level above it. Precipitation evaporates from the lake and the majority is recycled as precipitation before it can be carried away by the large-scale circulation. A very small portion escapes via this circulation and is cold-trapped on the night side, and integrated over sufficient time this limits the lifetime of the substellar lake. The lifetime itself depends sensitively on the initial surface moisture budget as well as on any horizontal moisture transport at ground (e.g., in a subsurface aquifer). The very slow depletion of the substellar lake suggests that only a small amount of surface moisture transport would be required to maintain the lake indefinitely. 
	\item Once the lake has evaporated, it appears unable to reform. The lifting condensation level of a dry lake region is sufficiently high that any moisture that reaches the substellar region is carried aloft by convection before the large-scale circulation whisks it away again. 
\end{enumerate}

These results have both similarities and differences to simulations with an aquaplanet. For example, \cite{labonte2020-Sensitivity} use an idealised GCM with a relatively similar configuration to this study, but a with a permanent moist surface and 30-day rotation rate. Their results display patterns of precipitation and evaporation similar to our results, with a positive surface moisture deficit ($P-E>0$) in a region east of the substellar point, and the remainder of the day side having a negative value. They suggest that equatorial convergence \citep{showman2011-EQUATORIAL} combines with the large-scale surface flow to return moisture from the night to the day side, producing precipitation over the substellar region. Assuming a land surface, this would produce a substellar lake of similar spatial extent to our results. \cite{ding2020-Stabilization} use a more complicated configuration using a bucket hydrology scheme and incorporating ice. They employ a different radiation profile based upon an M-type red dwarf and a 35-day rotation period, and an initial hydrology where all moisture starts on the night side as ice. Their experiments consider changes to the atmospheric \ch{CO2} concentration rather than the planetary moisture inventory. Critically, their simulations produce either an eyeball climate configuration or one with a substellar lake, depending upon \ch{CO2} concentration. They propose the large-scale interaction between high-altitude tropopause cold traps over the substellar point and the night side as a mechanism to allow a return flow of moisture to the substellar lake. This allows the lake to persist indefinitely, and to form when all moisture starts on the night side. In contrast, our results suggest a substellar lake can only form if surface liquid is present at the start of the tidally-locked state; even then, the surface layer must be sufficiently deep to sustain the lake for a significant period. Rather than a large-scale return flow sustaining the lake, we find that local effects, such as the height of the LCL, are the main control of the lake. In the absence of a (sub-)surface ground flow, we consider the substellar lake to be a transitional stage during the development of an eyeball climate, whose evolution is dependent upon the surface moisture budget. Rather than the tropopause cold traps, we suggest that the LCL is most important for controlling the substellar lake and climatic evolution our simulations. \citet{ding2021-Multiple} vary other planetary parameters using a grey-radiation scheme, but at no point is the moisture budget one of these. Their results indicate three different climate regimes: a night side ice cap; a substellar oasis (\textit{i.e. a lake}) and an intermediate substellar drizzle state. Our results share similarities with these states, but are distinguished by the fact that we consider the lake to be a transitional state. Although the lake is a transient feature in all our simulations, it's long lifetime in some simulations suggests that even a slow ground transport of water from nightside to day side could maintain the lake, although ice formation could completely prevent it. 

Our simulations have several limitations which should be taken into account. The absence of a cloud scheme \citep[e.g., that of][]{liu2021-SimCloud} means that precipitation occurs immediately upon saturation, potentially affecting atmospheric moisture transport and spatial distribution of moisture and precipitation. The relatively coarse resolution of T21 (needed for the very long integration times) may poorly resolve the small substellar lake and associated convective plumes in the overlying troposphere, limiting our ability to capture fine-scale dynamical processes that could influence the longevity of the lake \citep[e.g.,][]{sergeev2020-Atmospheric, sergeev2024-Impact}. The bucket scheme used lacks horizontal moisture transport preventing the modelling of surface runoff and/or subsurface flow that could return moisture to the day side; including that explicitly is needed for a quantitative study of this effect \citep{leconte2013-3D, faulk2020-Titans}. Ice formation (not present in our model) could, in some contrast, inhibit moisture transport back to the day side.   Still more generally, while the planets simulated in this work can be considered Earth-like, conditions including the hot, moist substellar point may push the boundaries of the various parameterizations schemes, in  particular the simple Betts--Miller scheme used for convection. Nevertheless, we believe that these results are of relevance with respect to understanding the diversity of planetary climates and the implications for habitability, particularly given the strong connection between total moisture inventory and surface features such as transient substellar lakes.

\begin{acknowledgments}
DAW was supported by an STFC studentship (ST/V506667/1). For the purpose of open access, the authors have applied a Creative Commons Attribution (CC BY) licence to any Author Accepted Manuscript version arising. Research data supporting this publication are openly available at [zenodo link to be updated before final submission]. The authors would like to thank [list to be updated before final submission].
\end{acknowledgments}

\bibliography{references}{}

@article{Adames_Wallace14,
   title =   {Three-dimensional structure and evolution of the {MJO} and its
relation to the mean flow},
   author =  {Adames, {\'A}ngel F. and Wallace, John M.},
   journal = {J. Atmos. Sci.},
   volume =  71,
   number =  6,
   pages =   {2007--2026},
   year =    2014,
}

@article{amundsen2016-UK,
  title = {The {{UK Met Office}} Global Circulation Model with a Sophisticated Radiation Scheme Applied to the Hot {{Jupiter HD}} 209458b},
  author = {Amundsen, David S. and Mayne, Nathan J. and Baraffe, Isabelle and Manners, James and Tremblin, Pascal and Drummond, Benjamin and Smith, Chris and Acreman, David M. and Homeier, Derek},
  year = 2016,
  month = oct,
  journal = {Astronomy and Astrophysics},
  volume = {595},
  pages = {A36},
  issn = {0004-6361},
  doi = {10.1051/0004-6361/201629183},
  urldate = {2023-01-17},
  langid = {english}
}

@article{barnes2017-Tidal,
  title = {Tidal Locking of Habitable Exoplanets},
  author = {Barnes, Rory},
  year = 2017,
  month = dec,
  journal = {Celestial Mechanics and Dynamical Astronomy},
  volume = {129},
  number = {4},
  pages = {509--536},
  issn = {1572-9478},
  doi = {10.1007/s10569-017-9783-7},
  urldate = {2025-01-24},
  langid = {english}
}

@article{betts1986-new,
  title = {A New Convective Adjustment Scheme. {{Part I}}: {{Observational}} and Theoretical Basis},
  shorttitle = {A New Convective Adjustment Scheme. {{Part I}}},
  author = {Betts, A. K.},
  year = 1986,
  journal = {Quarterly Journal of the Royal Meteorological Society},
  volume = {112},
  number = {473},
  pages = {677--691},
  issn = {1477-870X},
  doi = {10.1002/qj.49711247307},
  urldate = {2022-11-29},
  langid = {english}
}

@article{betts1986-newa,
  title = {A New Convective Adjustment Scheme. {{Part II}}: {{Single}} Column Tests Using {{GATE}} Wave, {{BOMEX}}, {{ATEX}} and Arctic Air-Mass Data Sets},
  shorttitle = {A New Convective Adjustment Scheme. {{Part II}}},
  author = {Betts, A. K. and Miller, M. J.},
  year = 1986,
  journal = {Quarterly Journal of the Royal Meteorological Society},
  volume = {112},
  number = {473},
  pages = {693--709},
  issn = {1477-870X},
  doi = {10.1002/qj.49711247308},
  urldate = {2022-11-29},
  langid = {english}
}

@article{checlair2017-No,
  title = {No {{Snowball}} on {{Habitable Tidally Locked Planets}}},
  author = {Checlair, Jade and Menou, Kristen and Abbot, Dorian S.},
  year = 2017,
  month = aug,
  journal = {The Astrophysical Journal},
  volume = {845},
  number = {2},
  pages = {132},
  publisher = {The American Astronomical Society},
  issn = {0004-637X},
  doi = {10.3847/1538-4357/aa80e1},
  urldate = {2025-01-31},
  langid = {english}
}

@article{cordier2021-Vertical,
  title = {Vertical Compositional Variations of Liquid Hydrocarbons in {{Titan}}'s Alkanofers},
  author = {Cordier, Daniel and Bonhommeau, David A. and Vu, Tuan H. and Choukroun, Mathieu and {Garc{\'i}a-S{\'a}nchez}, Fernando},
  year = 2021,
  month = sep,
  journal = {Astronomy \& Astrophysics},
  volume = {653},
  pages = {A80},
  publisher = {EDP Sciences},
  issn = {0004-6361, 1432-0746},
  doi = {10.1051/0004-6361/202140789},
  urldate = {2025-02-25},
  copyright = {{\copyright} D. Cordier et al. 2021},
  langid = {english}
}

@article{ding2020-Stabilization,
  title = {Stabilization of {{Dayside Surface Liquid Water}} via {{Tropopause Cold Trapping}} on {{Arid Slowly Rotating Tidally Locked Planets}}},
  author = {Ding, Feng and Wordsworth, Robin D.},
  year = 2020,
  month = mar,
  journal = {The Astrophysical Journal Letters},
  volume = {891},
  number = {1},
  pages = {L18},
  publisher = {The American Astronomical Society},
  issn = {2041-8205},
  doi = {10.3847/2041-8213/ab77d1},
  urldate = {2025-01-31},
  langid = {english}
}

@article{ding2021-Multiple,
  title = {Multiple {{Moist Climate Equilibrium States}} on {{Arid Rocky M-dwarf Planets}}: {{A Last-saturation Tracer Analysis}}},
  shorttitle = {Multiple {{Moist Climate Equilibrium States}} on {{Arid Rocky M-dwarf Planets}}},
  author = {Ding, Feng and Wordsworth, Robin D.},
  year = 2021,
  month = sep,
  journal = {The Planetary Science Journal},
  volume = {2},
  number = {5},
  pages = {201},
  publisher = {IOP Publishing},
  issn = {2632-3338},
  doi = {10.3847/PSJ/ac2236},
  urldate = {2025-02-01},
  langid = {english}
}

@article{edson2011-Atmospheric,
  title = {Atmospheric Circulations of Terrestrial Planets Orbiting Low-Mass Stars},
  author = {Edson, Adam and Lee, Sukyoung and Bannon, Peter and Kasting, James F. and Pollard, David},
  year = 2011,
  month = mar,
  journal = {Icarus},
  volume = {212},
  number = {1},
  pages = {1--13},
  issn = {0019-1035},
  doi = {10.1016/j.icarus.2010.11.023},
  urldate = {2024-10-29}
}

@article{edwards1996-Studies,
  title = {Studies with a Flexible New Radiation Code. {{I}}: {{Choosing}} a Configuration for a Large-Scale Model},
  shorttitle = {Studies with a Flexible New Radiation Code. {{I}}},
  author = {Edwards, J. M. and Slingo, A.},
  year = 1996,
  journal = {Quarterly Journal of the Royal Meteorological Society},
  volume = {122},
  number = {531},
  pages = {689--719},
  issn = {1477-870X},
  doi = {10.1002/qj.49712253107},
  urldate = {2023-01-17},
  langid = {english}
}

@article{faulk2020-Titans,
  title = {Titan's Climate Patterns and Surface Methane Distribution Due to the Coupling of Land Hydrology and Atmosphere},
  author = {Faulk, Sean P. and Lora, Juan M. and Mitchell, Jonathan L. and Milly, P. C. D.},
  year = 2020,
  month = apr,
  journal = {Nature Astronomy},
  volume = {4},
  number = {4},
  pages = {390--398},
  publisher = {Nature Publishing Group},
  issn = {2397-3366},
  doi = {10.1038/s41550-019-0963-0},
  urldate = {2021-08-11},
  copyright = {2019 The Author(s), under exclusive licence to Springer Nature Limited},
  langid = {english}
}

@article{frierson2007-Dynamics,
  title = {The {{Dynamics}} of {{Idealized Convection Schemes}} and {{Their Effect}} on the {{Zonally Averaged Tropical Circulation}}},
  author = {Frierson, Dargan M. W.},
  year = 2007,
  month = jun,
  journal = {Journal of the Atmospheric Sciences},
  volume = {64},
  number = {6},
  pages = {1959--1976},
  publisher = {American Meteorological Society},
  issn = {0022-4928, 1520-0469},
  doi = {10.1175/JAS3935.1},
  urldate = {2022-11-19},
  chapter = {Journal of the Atmospheric Sciences},
  langid = {english}
}

@article{genda2016-Origin,
  title = {Origin of {{Earth}}'s Oceans: {{An}} Assessment of the Total Amount, History and Supply of Water},
  shorttitle = {Origin of {{Earth}}'s Oceans},
  author = {Genda, Hidenori},
  year = 2016,
  journal = {Geochemical Journal},
  volume = {50},
  number = {1},
  pages = {27--42},
  doi = {10.2343/geochemj.2.0398}
}

@article{hammond2018-Wavemean,
  title = {Wave-Mean {{Flow Interactions}} in the {{Atmospheric Circulation}} of {{Tidally Locked Planets}}},
  author = {Hammond, Mark and Pierrehumbert, Raymond T.},
  year = 2018,
  month = dec,
  journal = {The Astrophysical Journal},
  volume = {869},
  number = {1},
  pages = {65},
  publisher = {The American Astronomical Society},
  issn = {0004-637X},
  doi = {10.3847/1538-4357/aaec03},
  urldate = {2024-11-08},
  langid = {english}
}

@article{haqq-misra2018-Demarcating,
  title = {Demarcating {{Circulation Regimes}} of {{Synchronously Rotating Terrestrial Planets}} within the {{Habitable Zone}}},
  author = {{Haqq-Misra}, Jacob and Wolf, Eric T. and Joshi, Manoj and Zhang, Xi and Kopparapu, Ravi Kumar},
  year = 2018,
  month = jan,
  journal = {The Astrophysical Journal},
  volume = {852},
  number = {2},
  pages = {67},
  publisher = {The American Astronomical Society},
  issn = {0004-637X},
  doi = {10.3847/1538-4357/aa9f1f},
  urldate = {2023-05-21},
  langid = {english}
}

@article{hayes2016-Lakes,
  title = {The {{Lakes}} and {{Seas}} of {{Titan}}},
  author = {Hayes, Alexander G.},
  year = 2016,
  month = jun,
  journal = {Annual Review of Earth and Planetary Sciences},
  volume = {44},
  number = {Volume 44, 2016},
  pages = {57--83},
  publisher = {Annual Reviews},
  issn = {0084-6597, 1545-4495},
  doi = {10.1146/annurev-earth-060115-012247},
  urldate = {2025-02-25},
  langid = {english}
}

@article{heller2011-Tidal,
  title = {Tidal Obliquity Evolution of Potentially Habitable Planets},
  author = {Heller, R. and Leconte, J. and Barnes, R.},
  year = 2011,
  month = apr,
  journal = {Astronomy \& Astrophysics},
  volume = {528},
  pages = {A27},
  publisher = {EDP Sciences},
  issn = {0004-6361, 1432-0746},
  doi = {10.1051/0004-6361/201015809},
  urldate = {2025-01-31},
  copyright = {{\copyright} ESO, 2011},
  langid = {english}
}

@article{joshi1997-Simulations,
  title = {Simulations of the {{Atmospheres}} of {{Synchronously Rotating Terrestrial Planets Orbiting M Dwarfs}}: {{Conditions}} for {{Atmospheric Collapse}} and the {{Implications}} for {{Habitability}}},
  shorttitle = {Simulations of the {{Atmospheres}} of {{Synchronously Rotating Terrestrial Planets Orbiting M Dwarfs}}},
  author = {Joshi, M. M. and Haberle, R. M. and Reynolds, R. T.},
  year = 1997,
  month = oct,
  journal = {Icarus},
  volume = {129},
  number = {2},
  pages = {450--465},
  issn = {0019-1035},
  doi = {10.1006/icar.1997.5793},
  urldate = {2023-05-23},
  langid = {english}
}

@article{kaspi2015-ATMOSPHERIC,
  title = {{{ATMOSPHERIC DYNAMICS OF TERRESTRIAL EXOPLANETS OVER A WIDE RANGE OF ORBITAL AND ATMOSPHERIC PARAMETERS}}},
  author = {Kaspi, Yohai and Showman, Adam P.},
  year = 2015,
  month = may,
  journal = {The Astrophysical Journal},
  volume = {804},
  number = {1},
  pages = {60},
  publisher = {American Astronomical Society},
  issn = {0004-637X},
  doi = {10.1088/0004-637X/804/1/60},
  urldate = {2022-07-20},
  langid = {english}
}

@article{koll2015-DECIPHERING,
  title = {{{DECIPHERING THERMAL PHASE CURVES OF DRY}}, {{TIDALLY LOCKED TERRESTRIAL PLANETS}}},
  author = {Koll, Daniel D. B. and Abbot, Dorian S.},
  year = 2015,
  month = mar,
  journal = {The Astrophysical Journal},
  volume = {802},
  number = {1},
  pages = {21},
  issn = {1538-4357},
  doi = {10.1088/0004-637X/802/1/21},
  urldate = {2025-01-07}
}

@article{koll2016-TEMPERATURE,
  title = {{{TEMPERATURE STRUCTURE AND ATMOSPHERIC CIRCULATION OF DRY TIDALLY LOCKED ROCKY EXOPLANETS}}},
  author = {Koll, Daniel D. B. and Abbot, Dorian S.},
  year = 2016,
  month = jul,
  journal = {The Astrophysical Journal},
  volume = {825},
  number = {2},
  pages = {99},
  publisher = {The American Astronomical Society},
  issn = {0004-637X},
  doi = {10.3847/0004-637X/825/2/99},
  urldate = {2025-07-23},
  langid = {english}
}

@article{komacek2019-Atmospheric,
  title = {The {{Atmospheric Circulation}} and {{Climate}} of {{Terrestrial Planets Orbiting Sun-like}} and {{M Dwarf Stars}} over a {{Broad Range}} of {{Planetary Parameters}}},
  author = {Komacek, Thaddeus D. and Abbot, Dorian S.},
  year = 2019,
  month = feb,
  journal = {The Astrophysical Journal},
  volume = {871},
  number = {2},
  pages = {245},
  publisher = {American Astronomical Society},
  issn = {0004-637X},
  doi = {10.3847/1538-4357/aafb33},
  urldate = {2022-10-03},
  langid = {english}
}

@article{labonte2020-Sensitivity,
  title = {Sensitivity of the {{Atmospheric Water Cycle}} within the {{Habitable Zone}} of a {{Tidally Locked}}, {{Earth-like Exoplanet}}},
  author = {Labont{\'e}, Marie-Pier and Merlis, Timothy M.},
  year = 2020,
  month = jun,
  journal = {The Astrophysical Journal},
  volume = {896},
  number = {1},
  pages = {31},
  publisher = {The American Astronomical Society},
  issn = {0004-637X},
  doi = {10.3847/1538-4357/ab9102},
  urldate = {2023-12-05},
  langid = {english}
}

@article{leconte2013-3D,
  title = {{{3D}} Climate Modeling of Close-in Land Planets: {{Circulation}} Patterns, Climate Moist Bistability, and Habitability},
  shorttitle = {{{3D}} Climate Modeling of Close-in Land Planets},
  author = {Leconte, J. and Forget, F. and Charnay, B. and Wordsworth, R. and Selsis, F. and Millour, E. and Spiga, A.},
  year = 2013,
  month = jun,
  journal = {Astronomy \& Astrophysics},
  volume = {554},
  pages = {A69},
  publisher = {EDP Sciences},
  issn = {0004-6361, 1432-0746},
  doi = {10.1051/0004-6361/201321042},
  urldate = {2023-04-17},
  copyright = {{\copyright} ESO, 2013},
  langid = {english}
}

@article{lewis2018-Influence,
  title = {The {{Influence}} of a {{Substellar Continent}} on the {{Climate}} of a {{Tidally Locked Exoplanet}}},
  author = {Lewis, Neil T. and Lambert, F. Hugo and Boutle, Ian A. and Mayne, Nathan J. and Manners, James and Acreman, David M.},
  year = 2018,
  month = feb,
  journal = {The Astrophysical Journal},
  volume = {854},
  number = {2},
  pages = {171},
  issn = {0004-637X, 1538-4357},
  doi = {10.3847/1538-4357/aaad0a},
  urldate = {2025-02-25}
}

@article{lewis2022-Temperature,
  title = {Temperature {{Structures Associated}} with {{Different Components}} of the {{Atmospheric Circulation}} on {{Tidally Locked Exoplanets}}},
  author = {Lewis, Neil T. and Hammond, Mark},
  year = 2022,
  month = dec,
  journal = {The Astrophysical Journal},
  volume = {941},
  number = {2},
  pages = {171},
  publisher = {The American Astronomical Society},
  issn = {0004-637X},
  doi = {10.3847/1538-4357/ac8fed},
  urldate = {2024-11-05},
  langid = {english}
}

@article{liu2021-SimCloud,
  title = {{{SimCloud}} Version 1.0: A Simple Diagnostic Cloud Scheme for Idealized Climate Models},
  shorttitle = {{{SimCloud}} Version 1.0},
  author = {Liu, Qun and Collins, Matthew and Maher, Penelope and Thomson, Stephen I. and Vallis, Geoffrey K.},
  year = 2021,
  month = may,
  journal = {Geoscientific Model Development},
  volume = {14},
  number = {5},
  pages = {2801--2826},
  publisher = {Copernicus GmbH},
  issn = {1991-959X},
  doi = {10.5194/gmd-14-2801-2021},
  urldate = {2024-11-05},
  langid = {english}
}

@article{manabe1969-CLIMATE,
  title = {{{CLIMATE AND THE OCEAN CIRCULATION}}},
  author = {Manabe, Syukuro},
  year = 1969,
  month = nov,
  journal = {Monthly Weather Review},
  issn = {1520-0493},
  urldate = {2024-09-13},
  chapter = {Monthly Weather Review},
  langid = {english}
}

@misc{manners2016-SOCRATES,
  title = {{{SOCRATES Technical Guide Suite Of Community RAdiative Transfer}} Codes Based on {{Edwards}} and {{Slingo}}},
  author = {Manners, James and Edwards, John M and Hill, Peter},
  year = 2016
}

@article{merlis2010-Atmospheric,
  title = {Atmospheric {{Dynamics}} of {{Earth-Like Tidally Locked Aquaplanets}}},
  author = {Merlis, Timothy M and Schneider, Tapio},
  year = 2010,
  journal = {Journal of Advances in Modeling Earth Systems},
  volume = {2},
  number = {4},
  issn = {1942-2466},
  doi = {10.3894/JAMES.2010.2.13},
  urldate = {2023-05-15},
  langid = {english}
}

@misc{nasaexoplanetscienceinstitute2020-Planetary,
  title = {Planetary {{Systems Table}}},
  author = {{NASA Exoplanet Science Institute}},
  year = 2020,
  publisher = {NExScI-Caltech/IPAC},
  doi = {10.26133/NEA12},
  urldate = {2023-05-15}
}

@article{noda2017-circulation,
  title = {The Circulation Pattern and Day-Night Heat Transport in the Atmosphere of a Synchronously Rotating Aquaplanet: {{Dependence}} on Planetary Rotation Rate},
  shorttitle = {The Circulation Pattern and Day-Night Heat Transport in the Atmosphere of a Synchronously Rotating Aquaplanet},
  author = {Noda, S. and Ishiwatari, M. and Nakajima, K. and Takahashi, Y. O. and Takehiro, S. and Onishi, M. and Hashimoto, G. L. and Kuramoto, K. and Hayashi, Y. -Y.},
  year = 2017,
  month = jan,
  journal = {Icarus},
  volume = {282},
  pages = {1--18},
  issn = {0019-1035},
  doi = {10.1016/j.icarus.2016.09.004},
  urldate = {2023-05-23},
  langid = {english}
}

@article{penn2017-Thermal,
  title = {The {{Thermal Phase Curve Offset}} on {{Tidally}} and {{Nontidally Locked Exoplanets}}: {{A Shallow Water Model}}},
  shorttitle = {The {{Thermal Phase Curve Offset}} on {{Tidally}} and {{Nontidally Locked Exoplanets}}},
  author = {Penn, James and Vallis, Geoffrey K},
  year = 2017,
  month = jun,
  journal = {The Astrophysical Journal},
  volume = {842},
  number = {2},
  pages = {101},
  issn = {0004-637X, 1538-4357},
  doi = {10.3847/1538-4357/aa756e},
  urldate = {2025-01-07}
}

@article{penn2018-Atmospheric,
  title = {Atmospheric {{Circulation}} and {{Thermal Phase-curve Offset}} of {{Tidally}} and {{Nontidally Locked Terrestrial Exoplanets}}},
  author = {Penn, James and Vallis, Geoffrey K.},
  year = 2018,
  month = dec,
  journal = {The Astrophysical Journal},
  volume = {868},
  number = {2},
  pages = {147},
  publisher = {The American Astronomical Society},
  issn = {0004-637X},
  doi = {10.3847/1538-4357/aaeb20},
  urldate = {2024-11-05},
  langid = {english}
}

@book{pierrehumbert2010-Principles,
  title = {Principles of {{Planetary Climate}}},
  author = {Pierrehumbert, Raymond T.},
  year = 2010,
  month = dec,
  publisher = {Cambridge University Press},
  googlebooks = {v8LXngEACAAJ},
  isbn = {978-0-521-86556-2},
  langid = {english}
}

@article{pierrehumbert2019-Atmospheric,
  title = {Atmospheric {{Circulation}} of {{Tide-Locked Exoplanets}}},
  author = {Pierrehumbert, Raymond T. and Hammond, Mark},
  year = 2019,
  journal = {Annual Review of Fluid Mechanics},
  volume = {51},
  number = {1},
  pages = {275--303},
  doi = {10.1146/annurev-fluid-010518-040516},
  urldate = {2023-08-08}
}

@article{salazar2020-Effect,
  title = {The {{Effect}} of {{Substellar Continent Size}} on {{Ocean Dynamics}} of {{Proxima Centauri}} b},
  author = {Salazar, Andrea M. and Olson, Stephanie L. and Komacek, Thaddeus D. and Stephens, Haynes and Abbot, Dorian S.},
  year = 2020,
  month = jun,
  journal = {The Astrophysical Journal Letters},
  volume = {896},
  number = {1},
  pages = {L16},
  publisher = {The American Astronomical Society},
  issn = {2041-8205},
  doi = {10.3847/2041-8213/ab94c1},
  urldate = {2025-07-23},
  langid = {english}
}

@article{sergeev2020-Atmospheric,
  title = {Atmospheric {{Convection Plays}} a {{Key Role}} in the {{Climate}} of {{Tidally Locked Terrestrial Exoplanets}}: {{Insights}} from {{High-resolution Simulations}}},
  shorttitle = {Atmospheric {{Convection Plays}} a {{Key Role}} in the {{Climate}} of {{Tidally Locked Terrestrial Exoplanets}}},
  author = {Sergeev, Denis E. and Lambert, F. Hugo and Mayne, Nathan J. and Boutle, Ian A. and Manners, James and Kohary, Krisztian},
  year = 2020,
  month = may,
  journal = {The Astrophysical Journal},
  volume = {894},
  number = {2},
  pages = {84},
  publisher = {The American Astronomical Society},
  issn = {0004-637X},
  doi = {10.3847/1538-4357/ab8882},
  urldate = {2023-02-08},
  langid = {english}
}

@article{sergeev2024-Impact,
  title = {The {{Impact}} of the {{Explicit Representation}} of {{Convection}} on the {{Climate}} of a {{Tidally Locked Planet}} in {{Global Stretched-mesh Simulations}}},
  author = {Sergeev, Denis E. and Boutle, Ian A. and Lambert, F. Hugo and Mayne, Nathan J. and Bendall, Thomas and Kohary, Krisztian and Olivier, Enrico and Shipway, Ben},
  year = 2024,
  month = jul,
  journal = {The Astrophysical Journal},
  volume = {970},
  number = {1},
  pages = {7},
  publisher = {The American Astronomical Society},
  issn = {0004-637X},
  doi = {10.3847/1538-4357/ad4ecd},
  urldate = {2025-01-30},
  langid = {english}
}

@article{showman2002-Atmospheric,
  title = {Atmospheric Circulation and Tides of ``51 {{Pegasus}} b-like'' Planets},
  author = {Showman, A. P. and Guillot, T.},
  year = 2002,
  month = apr,
  journal = {Astronomy \& Astrophysics},
  volume = {385},
  number = {1},
  pages = {166--180},
  publisher = {EDP Sciences},
  issn = {0004-6361, 1432-0746},
  doi = {10.1051/0004-6361:20020101},
  urldate = {2024-08-01},
  copyright = {{\copyright} ESO, 2002},
  langid = {english}
}

@article{showman2011-EQUATORIAL,
  title = {{{EQUATORIAL SUPERROTATION ON TIDALLY LOCKED EXOPLANETS}}},
  author = {Showman, Adam P. and Polvani, Lorenzo M.},
  year = 2011,
  month = aug,
  journal = {The Astrophysical Journal},
  volume = {738},
  number = {1},
  pages = {71},
  publisher = {The American Astronomical Society},
  issn = {0004-637X},
  doi = {10.1088/0004-637X/738/1/71},
  urldate = {2024-10-29},
  langid = {english}
}

@article{song2021-Asymmetry,
  title = {Asymmetry and {{Variability}} in the {{Transmission Spectra}} of {{Tidally Locked Habitable Planets}}},
  author = {Song, Xinyi and Yang, Jun},
  year = 2021,
  month = sep,
  journal = {Frontiers in Astronomy and Space Sciences},
  volume = {8},
  pages = {708023},
  issn = {2296-987X},
  doi = {10.3389/fspas.2021.708023},
  urldate = {2025-01-07}
}

@article{thomson2019-Hierarchical,
  title = {Hierarchical {{Modeling}} of {{Solar System Planets}} with {{Isca}}},
  author = {Thomson, Stephen I and Vallis, Geoffrey K},
  year = 2019,
  month = dec,
  journal = {Atmosphere},
  volume = {10},
  number = {803},
  pages = {21},
  langid = {english}
}

@article{Vallis21,
   author =  {G. K. Vallis},
   title =   {Distilling the mechanism of the {Madden--Julian} Oscillation to a simple translating structure},
   journal = {Quart. J. Roy. Meteor. Soc.},
   year =    2021,
   doi = {10.1002/qj.4114},
   volume = {},
   pages =   {1--16},
}

@article{Vallis_Penn20,
  title={Convective Organization and Eastward Propagating Equatorial Disturbances in a Simple Excitable System},
  author={Vallis, Geoffrey K and Penn, James},
  journal={Quart. J. Roy. Meteor. Soc.},
  year={2020},
  pages={1-18},
  doi = {10.1002/QJ.3792},
  note = {{doi}:10.1002/QJ.3792},
}

@article{vallis2018-Isca,
  title = {Isca, v1.0: A Framework for the Global Modelling of the Atmospheres of {{Earth}} and Other Planets at Varying Levels of Complexity},
  shorttitle = {Isca, v1.0},
  author = {Vallis, Geoffrey K. and Colyer, Greg and Geen, Ruth and Gerber, Edwin and Jucker, Martin and Maher, Penelope and Paterson, Alexander and Pietschnig, Marianne and Penn, James and Thomson, Stephen I.},
  year = 2018,
  month = mar,
  journal = {Geoscientific Model Development},
  volume = {11},
  number = {3},
  pages = {843--859},
  issn = {1991-9603},
  doi = {10.5194/gmd-11-843-2018},
  urldate = {2021-04-20},
  langid = {english}
}

@article{way2017-Resolving,
  title = {Resolving {{Orbital}} and {{Climate Keys}} of {{Earth}} and {{Extraterrestrial Environments}} with {{Dynamics}} ({{ROCKE-3D}}) 1.0: {{A General Circulation Model}} for {{Simulating}} the {{Climates}} of {{Rocky Planets}}},
  shorttitle = {Resolving {{Orbital}} and {{Climate Keys}} of {{Earth}} and {{Extraterrestrial Environments}} with {{Dynamics}} ({{ROCKE-3D}}) 1.0},
  author = {Way, M. J. and Aleinov, I. and Amundsen, David S. and Chandler, M. A. and Clune, T. L. and Genio, A. D. Del and Fujii, Y. and Kelley, M. and Kiang, N. Y. and Sohl, L. and Tsigaridis, K.},
  year = 2017,
  month = jul,
  journal = {The Astrophysical Journal Supplement Series},
  volume = {231},
  number = {1},
  pages = {12},
  publisher = {The American Astronomical Society},
  issn = {0067-0049},
  doi = {10.3847/1538-4365/aa7a06},
  urldate = {2023-01-17},
  langid = {english}
}

@article{witte2002-Orbital,
  title = {Orbital Evolution by Dynamical Tides in Solar Type Stars - {{Application}} to Binary Stars and Planetary Orbits},
  author = {Witte, M. G. and Savonije, G. J.},
  year = 2002,
  month = apr,
  journal = {Astronomy \& Astrophysics},
  volume = {386},
  number = {1},
  pages = {222--236},
  publisher = {EDP Sciences},
  issn = {0004-6361, 1432-0746},
  doi = {10.1051/0004-6361:20020155},
  urldate = {2025-01-24},
  copyright = {{\copyright} ESO, 2002},
  langid = {english}
}

@article{wordsworth2022-Atmospheres,
  title = {Atmospheres of {{Rocky Exoplanets}}},
  author = {Wordsworth, Robin and Kreidberg, Laura},
  year = 2022,
  journal = {Annual Review of Astronomy and Astrophysics},
  volume = {60},
  number = {1},
  pages = {159--201},
  doi = {10.1146/annurev-astro-052920-125632},
  urldate = {2023-02-08}
}

@article{yang2013-STABILIZING,
  title = {{{STABILIZING CLOUD FEEDBACK DRAMATICALLY EXPANDS THE HABITABLE ZONE OF TIDALLY LOCKED PLANETS}}},
  author = {Yang, Jun and Cowan, Nicolas B. and Abbot, Dorian S.},
  year = 2013,
  month = jun,
  journal = {The Astrophysical Journal Letters},
  volume = {771},
  number = {2},
  pages = {L45},
  publisher = {The American Astronomical Society},
  issn = {2041-8205},
  doi = {10.1088/2041-8205/771/2/L45},
  urldate = {2023-02-08},
  langid = {english}
}

@article{yang2019-Simulations,
  title = {Simulations of {{Water Vapor}} and {{Clouds}} on {{Rapidly Rotating}} and {{Tidally Locked Planets}}: {{A 3D Model Intercomparison}}},
  shorttitle = {Simulations of {{Water Vapor}} and {{Clouds}} on {{Rapidly Rotating}} and {{Tidally Locked Planets}}},
  author = {Yang, Jun and Leconte, J{\'e}r{\'e}my and Wolf, Eric T. and Merlis, Timothy and Koll, Daniel D. B. and Forget, Fran{\c c}ois and Abbot, Dorian S.},
  year = 2019,
  month = apr,
  journal = {The Astrophysical Journal},
  volume = {875},
  number = {1},
  pages = {46},
  publisher = {The American Astronomical Society},
  issn = {0004-637X},
  doi = {10.3847/1538-4357/ab09f1},
  urldate = {2023-02-08},
  langid = {english}
}
\bibliographystyle{aasjournal}

\end{document}